\documentclass[traditabstract]{aa}

\usepackage{amsmath}
\usepackage{natbib}
\usepackage{graphicx}
\usepackage{xspace}
\usepackage{float}
\usepackage{hyperref}
\usepackage{color}

\definecolor{darkblue}{rgb}{0.0,0.0,0.7}

\hypersetup{
colorlinks=true,
linkcolor=darkblue,
anchorcolor=darkblue,
citecolor=darkblue,	
filecolor=darkblue,
menucolor=darkblue,
runcolor	=darkblue, 	
urlcolor=darkblue
}

\bibpunct{(}{)}{;}{a}{}{,}

\newcommand{\hyperion}{\textsc{Hyperion}\xspace}

\begin{document}

\title{HYPERION: An open-source parallelized three-dimensional dust continuum radiative transfer code}

\author{Thomas P. Robitaille\inst{1,2}}

\institute{Harvard-Smithsonian Center for Astrophysics, 60 Garden Street, Cambridge, MA, 02138, USA \\
           \email{trobitaille@cfa.harvard.edu}
           \and
           \textit{Spitzer} Postdoctoral Fellow
          }

\date{Received 27 April 2011, Accepted 28 September 2011}

\abstract{
\hyperion is a new three-dimensional dust continuum Monte-Carlo radiative transfer code that is designed to be as generic as possible, allowing radiative transfer to be computed through a variety of three-dimensional grids. The main part of the code is problem-independent, and only requires an arbitrary three-dimensional density structure, dust properties, the position and properties of the illuminating sources, and parameters controlling the running and output of the code. \hyperion is parallelized, and is shown to scale well to thousands of processes. Two common benchmark models for protoplanetary disks were computed, and the results are found to be in excellent agreement with those from other codes. Finally, to demonstrate the capabilities of the code, dust temperatures, SEDs, and synthetic multi-wavelength images were computed for a dynamical simulation of a low-mass star formation region. \hyperion is being actively developed to include new features, and is publicly available (\url{http://www.hyperion-rt.org}).
}{}


\titlerunning{\hyperion}
\authorrunning{Robitaille}

\maketitle

\section{Introduction}

\label{sec:introduction}

The investigation of astrophysical sources via multi-wavelength observations requires understanding the transfer of radiation through dust and gas in order to reliably derive geometrical, physical, and chemical properties. While a small subset of problems can be solved analytically, most require a numerical solution.

A variety of techniques have been developed to this effect, and until recently, most relied on a direct numerical solution to the differential equation of radiative transfer. Examples include 2$^{\rm nd}$ order finite differences \citep{steinacker:03:405}, 5$^{\rm th}$ order Runge-Kutta integration \citep{steinacker:06:920}, and variable Eddington tensors \citep{dullemond:00:1187, dullemond:02:853}.

While these techniques can often provide results accurate to arbitrary precision, and have been very successful for one-dimensional problems, they become increasingly complex when applied to two- and three-dimensional problems. Within the last decade, the \textit{Monte-Carlo} technique applied to radiative transfer has become an increasingly popular alternative that is well suited to arbitrary three-dimensional density distributions. Rather than solving the equation of radiative transfer directly, this method relies on random sampling of probability distribution functions to propagate \textit{photon packets} through a grid of constant density and temperature cells.

In its simplest implementation, Monte-Carlo radiative transfer considers single-frequency photon packets that are emitted by sources, and are propagated through a grid of constant density cells. Each photon packet travels a certain optical depth, before being either scattered, or absorbed and immediately re-emitted (to conserve energy). These photon packets can then continue to travel and interact until they escape the grid. The optical depth, type of interaction, and frequency of re-emitted photon packets, as well as scattering/re-emission directions are all sampled from probability distribution functions. Photon packets escaping from the grid can be used to compute synthetic observations, such as spectral energy distributions (SEDs) and images. By repeating this process for large numbers of photon packets, the signal-to-noise of the synthetic observations can be increased.

In cases where local thermodynamic equilibrium (LTE) can be assumed, the energy absorbed in each cell can be used to compute the equilibrium temperature. Since the emissivity depends on the temperature, the radiative transfer has to be run iteratively to obtain a self-consistent solution for the temperature.

Various methods have been developed to improve the performance of Monte-Carlo radiative transfer codes, which would otherwise be inefficient in many cases.
For example, rather than simply using discrete absorptions in cells to compute the equilibrium temperature, \citet{Lucy:99:282} suggested summing the optical depth to absorption of photon packet paths through each grid cell, requiring far fewer photons to achieve a comparable signal-to-noise in the derived temperatures. \citet{bjorkman:01:615} developed an algorithm whereby the temperature of a cell is updated immediately each time a photon is absorbed, and the frequency of the re-emitted photon is sampled from a difference emissivity function that introduces a correction to account for the fact that previous photons were sampled from emissivities for different temperatures. This algorithm is sometimes referred to simply as `immediate re-emission', but this is a misnomer since other algorithms also immediately re-emit a photon following an absorption to conserve energy \citep[e.g.][]{Lucy:99:282}; instead, what this algorithm introduces is the \textit{immediate temperature correction} aspect.

For synthetic observations, the \textit{peeling-off} technique \citep{Yusef-Zadeh:84:186} greatly improves the signal-to-noise of SEDs and images: every time a photon packet is emitted, scattered, or re-emitted after an absorption, the probability that the emitted or scattered photon would reach the observer is used to build up the observations. Thus each photon contributes multiple times to the SEDs or images. Another example is that of \textit{forced first scattering} \citep[e.g.][]{mattila:70:53, wood:99:799}, which can be used in optically thin cases to force photon packets to interact with the dust (this requires weighting the photons accordingly to avoid any bias due to the forcing).

In standard Monte-Carlo, the number of interactions a photon packet will undergo before traveling a certain distance increases approximately with the square of the density, so that in very optically thick cells, photon packets may become effectively trapped. This problem can be avoided by locally making use of the diffusion approximation to solve the radiative transfer in these regions. For example, \citet{min:09:155} developed a modified random walk procedure based on the diffusion approximation that can dramatically reduce the number of steps required for a photon to escape from a grid cell of arbitrarily large optical depth.

A large collection of dust continuum Monte-Carlo radiative transfer codes has been developed  \citep[e.g.][]{wolf:99:839, harries:00:722, gordon:01:269, misselt:01:277, wood:01:299, wood:02:1183, wood:02:887, wolf:03:99, stamatellos:03:941, whitney:03:1079, whitney:03:1049, whitney:04:1177, dullemond:04:159, jonsson:06:2, pinte:06:797, min:09:155}, each including some or all of the above optimizations, as well as other optimizations not mentioned here. While some of the early codes assumed spherical or axis-symmetric geometries for simplicity, many have since been adapted to compute fully arbitrary three-dimensional distributions. In addition to dust continuum radiative transfer, some codes can also compute non-LTE line transfer \citep{carciofi:06:1081, carciofi:08:1374}, or photoionization \citep[e.g.][]{ercolano:03:1136, ercolano:05:1038, ercolano:08:534}.

This paper presents \hyperion, a new dust-continuum Monte-Carlo radiative transfer code that is designed to be applicable to a wide range of problems. \hyperion implements many of the recent optimizations to the Monte-Carlo technique discussed above, was written from the start to be a parallel code that can scale to thousands of processes, and is written in a modular and extensible way so as to be easily improved in future. It can treat the emission from an arbitrary number of sources, can include multiple dust types, and can compute the anisotropic scattering of polarized radiation using fully numerical scattering phase functions. It uses the \citet{Lucy:99:282} iterative method to determine the radiative equilibrium temperature, but does not use the \citet{bjorkman:01:615} temperature correction technique, as the latter is much more challenging to parallelize efficiently. Thanks to the modular nature of the code, the radiative transfer can be computed on a number of different three-dimensional grid types, and additional grid types can be added in future. \hyperion can compute SEDs and multi-wavelength images and polarization maps. The code is released under an open source license, and is hosted on a service that allows members of the community to easily contribute to the code and documentation.

Section \ref{sec:overview} gives an overview of the implementation of the code. Section \ref{sec:parallel} discusses the efficiency of the parallelized code. Section \ref{sec:benchmarks} presents the results for two benchmark models of protoplanetary disks. Finally, Section \ref{sec:simulation} demonstrates the capabilities of the code by computing temperatures, SEDs, and synthetic images for a simulation of a star-formation region. The availability of the code and plans for the future are discussed in Section \ref{sec:future}.

\section{Code overview}

\label{sec:overview}

The code is split into two main components. The first, which carries out the core of the radiative transfer calculation, is implemented in Fortran 95/2003 for high performance. This part of the code is problem-independent: the input (bundled into a single file) consists of an arbitrary three-dimensional density structure as well as dust properties, a list of sources, and output parameters. This input is used by the Monte-Carlo radiative transfer code to compute temperatures, SEDs, and images.  Therefore, it is possible to use either gridded analytical density structures, or arbitrary density grids from simulations. At the moment, \hyperion supports several types of three-dimensional grids (\S\ref{sec:grid}) and the modular nature of the code will make it easy to add support for additional grid types in the future. It is possible to specify an arbitrary number of dust types (within computational limits), which allows models to have different effective grain size distributions and compositions in different grid cells.

The second component of the code consists of an object-oriented Python library that makes it easy to set up the input file for arbitrary problems from a single script. This library includes pre-defined analytical density structures for common problems such as flared disks and rotationally flattened envelopes and will also include scripts to import density structures from simulations. Post-processing tools are also provided in the Python library to analyze the results of radiative transfer models.

The present section describes the algorithm for the main radiative transfer code. The code first reads in the inputs (\S\ref{sec:input}), then propagates photon packets through the grid (\S\ref{sec:propagation}) for multiple iterations to compute the energy absorbed in each cell (\S\ref{sec:temperature}). Once the absorbed energy calculation has converged (\S\ref{sec:convergence}), the code computes SEDs and images (\S\ref{sec:seds}).

\subsection{Inputs to the code}

\label{sec:input}

\subsubsection{Sources}

A model can include any number of sources of emission (within computational limits). Each source is characterized by a bolometric luminosity, and the frequencies of the emitted photon packets are randomly sampled such that the emergent frequency distribution of the packets reproduces a user-defined spectrum. The total number of photons to emit from sources is set by the user. A number of different source types can be used -- at the moment, the code supports:
\begin{itemize}
\item Isotropic point sources.
\item Spherical sources with or without limb darkening. These can include arbitrary numbers of cool or hot spots, each with different positions, sizes, and with their own spectrum.
\item Diffuse sources where flux is emitted from within grid cells according to a three-dimensional probability distribution function (this can be useful for unresolved stellar populations in galaxy models, or for accretion luminosity emitted via viscous energy dissipation in protoplanetary disks).
\item External isotropic sources, which can be used to simulate an interstellar or intergalactic radiation field.
\end{itemize}
Each photon packet emitted is characterized by a position, direction vector, frequency, and a Stokes vector (I, Q, U, V) that describes the total intensity and the linear and circular polarization.

\subsubsection{Dust density grid}

\label{sec:grid}

The code is written in a modular way that allows support for different grid geometries to be easily added. At the moment, three-dimensional cartesian, spherical-polar, and cylindrical-polar grids can be used, as well as two types of adaptive cartesian grids. The first is a standard \textit{octree} grid, in which each cubic cell can be recursively split equally into eight smaller cubic cells. The second is the type of grid commonly used in adaptive mesh refinement (AMR) hydrodynamical codes. Here, a coarse grid is first defined on the zero-th level of refinement, and with increasing levels, increasingly finer grids can be used in  areas where high resolution is needed. Because they concentrate the resolution where it is needed, variable resolution grids such as octrees and AMR allow radiation transfer to be computed on density grids with effective resolutions that would be prohibitive with regular cartesian grids.

\subsubsection{Dust properties}

\label{sec:dust}

The dust properties required are the frequency-dependent mass extinction coefficient $\chi_\nu$ and albedo $\omega_\nu$, as well as the scattering properties of the dust. At this time, \hyperion supports anisotropic wavelength-dependent scattering of randomly oriented grains, using a 4-element Mueller matrix \citep{Chandrasekhar:60, Code:95:400}:
\begin{equation}
\left(\begin{array}{c}I \\Q \\U \\V\end{array}\right)_{\rm scattered} = \left(\begin{array}{cccc}S_{11} & S_{12} & 0 & 0 \\S_{12} & S_{11} & 0 & 0 \\0 & 0 & S_{33} & -S_{34} \\0 & 0 & S_{34} & S_{33}\end{array}\right) \cdot \left(\begin{array}{c}I \\Q \\U \\V\end{array}\right)_{\rm incident}
\end{equation}
Support for aligned non-spherical dust grains, which are described by a full 16-element matrix, will be implemented in future \citep[e.g.][]{whitney:02:205}.

To keep the Fortran code as general as possible, the mean opacities and emissivities of the dust are pre-computed by the Python library as a function of the specific energy absorption rate of the dust rather than the dust temperature (c.f. \S\ref{sec:temperature}). For dust in LTE, the emissivities are given by $j_\nu = \kappa_\nu\,B_\nu(T)$, and the mean opacities are the usual Planck and Rosseland mean opacities to extinction and absorption. However, it is also possible to specify mean opacities and emissivities that do not assume LTE (e.g. \S\ref{sec:pah}). Thus, assumptions about LTE are made at the level of the dust files, rather than in the radiative transfer code itself.

\subsection{Photon packet propagation}

\label{sec:propagation}

The code implements the propagation of photon packets in the following way: a photon packet is emitted from a source, randomly selected from a probability distribution function defined by the relative luminosity of the different sources. This sampling can be done either in the standard way to give a number of photon packets proportional to the source luminosity, or to give equal numbers of photons to each source, which requires weighting the energy of the photons. The direction and frequency of the photon packet are randomly sampled accordingly for the type and the spectrum of the source it originates from, using standard sampling with no weighting. A random optical depth to extinction $\tau$ is sampled from the probability distribution function $\exp{(-\tau)}$ by sampling a random number $\zeta$ uniformly between 0 and 1, and taking $\tau=-\ln{\zeta}$. The photon packet is then propagated along a ray until it either escapes the grid, or reaches the sampled optical depth, whichever happens first. If the photon packet has not escaped the grid, it will then interact with the dust. A random number $\zeta$ is sampled uniformly between 0 and 1, and if it is larger than the albedo of the dust, the photon packet is absorbed; otherwise it is scattered. Once the photon packet is scattered or re-emitted, a new optical depth is sampled, and the process is repeated until the photon packet escapes from the grid.

Very optically thick regions are an issue in basic Monte-Carlo radiative transfer, as photon packets can get trapped in these regions and require in some cases millions of absorptions/re-emissions and scatterings to escape. Recently, \cite{min:09:155} proposed a modified random walk (MRW) algorithm that allows photon packets to propagate efficiently in very optically thick regions by grouping many scatterings and absorptions/re-emissions into single larger steps. The photon packet propagation described previously is done in a grid made up of cells of constant density and temperature. Therefore if the mean optical depth to the edge of a cell is much larger than unity, one can set up a sphere whose radius is smaller than the distance to the closest wall, inside which the density will be constant, and travel to the edge of a sphere in a single step using the diffusion approximation, thus avoiding the need to compute millions of interactions. \hyperion includes the implementation of the MRW algorithm described in \citet{robitaille:10:A70}.

\subsection{Temperature/Energy absorption rate calculation}

\label{sec:temperature}

\hyperion uses the iterative continuous absorption method proposed by \citet{Lucy:99:282}: in the first iteration, the specific energy absorption rate of the dust is computed in each cell using
\begin{equation}
\dot{A} = \frac{1}{\Delta t}\,\frac{\epsilon}{V}\,\sum \ell\,\kappa_\nu
\end{equation}
where $\Delta t$ is the time over which photon packets are emitted (taken to be 1\,s in the code), $V$ is the cell volume, $\epsilon$ is the energy of a photon packet, $\ell$ is the path length traveled -- which depends on the density, as does the \textit{number} of path lengths being added -- and $\kappa_\nu$ is the mass absorption coefficient. The temperature $T$ of the dust can be found from $\dot{A}$ by balancing absorbed and emitted energy, assuming LTE:
\begin{equation}
4\pi\,\kappa_P(T)\,B(T) = \dot{A}
\end{equation}
where $\kappa_P(T)$ is the Planck mean mass absorption coefficient, and $B(T)=(\sigma/\pi)\,T^4$ is the integral of the Planck function. As mentioned in \S\ref{sec:dust}, \hyperion does not compute temperatures, but instead the mean opacities and emissivities of the dust, which are pre-computed, are tabulated as a function of $\dot{A}$ rather than $T$. In cases where the temperature is needed (for example if requested as output from the user), the temperature is computed on the fly using the pre-computed Planck mean opacities.

For high optical depth problems, such as protoplanetary disks, some regions in the grid may see few or no photon packets, and reliable values for $\dot{A}$ or $T$ can therefore not be directly computed. In this case, one can formally solve the diffusion approximation for the cells that see fewer than a given threshold of photon packets, using cells that do have reliable values as boundary conditions (see Appendix \ref{app:pda}). This is referred to as the partial diffusion approximation (PDA) in \citet{min:09:155}.

\subsection{Convergence}

\label{sec:convergence}

The emissivity of the dust depends on $\dot{A}$, which depends on the propagation of the photon packets and the frequency of the photon packets, which in turn depends on the emissivity of the dust, so one needs to compute the radiative transfer for several iterations before the values of $\dot{A}$ in each cell converge. A simple algorithm to determine convergence is included in the code. The main function used in the convergence algorithm is
\begin{equation}
\delta(x_1, x_2) = {\rm max}\left(\frac{x_1}{x_2}, \frac{x_2}{x_1}\right)
\end{equation}
This measures how different $x_2$ is from $x_1$ and does not depend on the direction of the change. For example, a value of $\delta=2$ means that the quantity has changed by a factor of 2. Because the change is expressed as a ratio, this means that large changes can be more easily expressed than using a simple fractional difference.

At each iteration $i$, the code determines by how much the energy absorbed $\dot{A}$ has changed by computing for each cell $j$ the change in the energy absorbed in each cell $j$,
\begin{equation}
R_{j}^{i} \equiv \delta(\dot{A}_{j}^{i-1}, \dot{A}_{j}^{i})
\end{equation}
The quantile value $Q^i$ at the $p$-th percentile of the $R_j$ values is then computed and compared to the value found during the previous iteration, $Q^{i-1}$. Finally, the change in this quantile is calculated using
\begin{equation}
\Delta^i \equiv \delta\left(Q^{i-1}, Q^i\right)
\end{equation}
The specific energy absorption rate is considered to be converged once $Q^i$ and $\Delta^i$ have fallen below user specified values $Q_{\rm thres}$ and $\Delta_{\rm thres}$. As an example, setting $p=99.9\%$, $Q_{\rm thres}=2.$, and $\Delta_{\rm thres}=1.1$ means that convergence is achieved once 99.9\% of the differences in specific energy absorption rates between iterations are less than a factor of 2, and once the 99.9\% percentile value of the difference changes by less than a factor of 1.1 ($\sim10$\%).

Using this techniques is more robust to noise in the specific energy values than simply requiring that \textit{all} cells vary less than a given threshold, since it allows the user to ignore outliers by setting the percentile value appropriately. Of course, there is no guarantee that the criteria set by the user will be met at any iteration if the noise is too large, but in this case the user will see that there is not convergence, and can increase the number of photon packets, the maximum number of iterations, or use less stringent requirements for convergence. In any case, this method allows users to set quantitative requirements that are necessary to meet their scientific problem and that are more flexible than a single threshold on all values.

\subsection{PAH/VSG emission}

\label{sec:pah}

In addition to dust continuum radiative transfer, the code can also take into account any population where the opacity is independent of temperature or density, and for which the emissivity depends only on the specific energy absorbed ($\dot{A}$) inside each cell (\S\ref{sec:dust}). While this does not permit arbitrary non-LTE radiative transfer, it does allow one to use an approximation of the emission from stochastically heated polycyclic aromatic hydrocarbons (PAHs) and very small non-thermal grains (VSGs) using a similar prescription to that given in \citet{wood:08:1118}.

The idea behind the algorithm is to pre-compute the average emissivity of an ensemble of PAHs and VSGs, as a function of the energy deposited by radiation into the PAHs and VSGs -- the specific energy absorption rate $\dot{A}$ for the PAH and VSG populations -- for a given irradiating spectrum, and to then use these emissivities as look-up tables in the radiative transfer code.

Since the opacities of PAHs and VSGs are typically strongly peaked in the UV and optical, this means that the UV and optical will dominate the energy exciting and being reprocessed by the PAHs and VSGs, and the emissivities are then being chosen based on the total strength of the UV and optical emission relative to the original template spectrum. The \textit{shape} of the template spectrum does have a small impact on the pre-computed emissivities, but to first order, simply using the ratio of the total energy absorbed should be adequate for estimating the importance of PAHs and VSGs as absorbers and emitters, and allows the impact on the SEDs and images to be studied. While the \textit{strength} of the PAH features and VSG continuum in SEDs should be accurate to first order, the \textit{shape} of the PAH features should be treated with caution.

Note that the implementation discussed here differs in one respect from the \citet{wood:08:1118} prescription: the latter uses the mean intensity, rather than the energy intercepted by the PAHs and VSGs, to determine which emissivity file to use. Using the energy absorbed by the PAHs and VSGs may be more appropriate than using the mean intensity, since the latter would predict the same excitation for a fixed mean intensity, whether this intensity peaked in the UV or the millimeter, while using $\dot{A}$ would give a higher excitation for a spectrum peaking in the UV compared to one peaking in the mm.

\subsection{SEDs and images}

\label{sec:seds}

Once the specific energy absorption rate calculation has converged, SEDs and images can be computed. There are several methods to do this, and all of the following are implemented in the code.

\subsubsection{Photon binning}

\label{sec:binning}

The easiest and most inefficient method to compute SEDs and images is to propagate the photon packets as for the initial iterations, and bin them all into viewing angles as they escape from the grid. This is very inefficient, because each photon packet only contributes once to the SEDs and images, and only to one viewing angle. Furthermore, the viewing angle bins cannot be arbitrarily small, and therefore the SEDs and images resulting from this are averaged over viewing angle.

\subsubsection{Peeling-off}

\label{sec:peelingoff}

\cite{Yusef-Zadeh:84:186} introduced the concept of peeling-off, whereby at each scattering or re-emission, the probability $p$ of the photon packet being scattered or re-emitted towards the observer immediately after the interaction is computed, and a photon packet with weight $p\,\exp{(-\tau)}$ is added to the SED and images, where $\tau$ is the optical depth to reach the observer from the interaction. This results in much higher signal-to-noise than photon packet binning, because each photon packet contributes several times to the SEDs and images at each viewing angle.

\subsubsection{Raytracing}

\label{sec:raytracing}

By far the most efficient method of computing SEDs and images is raytracing, which essentially consists of determining the source function at each position in the grid, and solving -- in post-processing -- the equation of radiative transfer along lines of sight to the observer through the dust geometry. For thermal emission, this is relatively straightforward, because the source function in each cell is simply related to the mass and temperature or energy in the cell. For scattered light, unless the scattering phase function is isotropic, one needs to retain information about the angular and frequency dependence of the incident or scattered light. As discussed in \cite{Pinte:09:967}, this is either computationally very expensive in terms of memory, or results in a loss of accuracy for strongly peaked scattering phase functions. In the current implementation of \hyperion, raytracing can be used for the source and dust emission, allowing excellent signal-to-noise to be achieved at long wavelengths where traditional Monte-Carlo only produces very few photon packets. Raytracing for scattered light will be implemented as an option in future since -- if adequate computational resources are available -- it can provide excellent signal-to-noise significantly faster than conventional peeling-off as for source and dust emission.

\subsubsection{Monochromatic radiative transfer}

The default mode of the binning (\S\ref{sec:binning}), peeling-off (\S\ref{sec:peelingoff}), and raytracing (\S\ref{sec:raytracing}) algorithms is to produce SEDs and images with finite width wavelength/frequency bins. However, in some cases it is desirable to compute SEDs or images at exact wavelengths/frequencies. When this is the case, the peeling-off algorithm for computing the scattered light contribution to the SED or images has to be modified. In this case, the scattered light is separated into two contributions, namely the  scattered light from the sources, and the scattered dust emission. To compute the scattered light from the sources, photon packets are emitted by all the sources at the fixed wavelengths/frequencies required. Each time a photon packet scatters, a photon packet is peeled-off, and each time a photon packet is absorbed, it is terminated. Thus, there is no immediate re-emission following an absorption. The next step is to compute the scattered dust emission. The specific energy absorption rate is used to compute the emissivity at the fixed wavelengths/frequencies inside each cell. To emit a photon packet, a random cell is selected in the grid, and the photon packet is emitted randomly within the cell, carrying an amount of energy proportional to the local emissivity. The photon packet is then propagated, and a photon packet is peeled-off at each scattering. As before, the photon packet is terminated once it is absorbed. This algorithm ensures conservation of the total scattered light contribution.

\begin{figure*}
\includegraphics[width=0.98\textwidth]{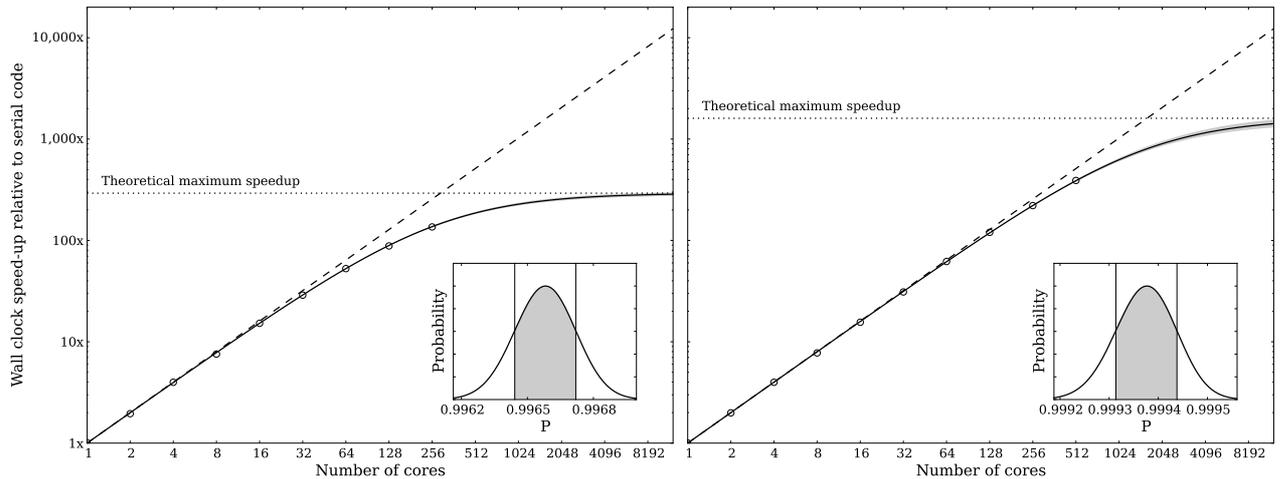}
\caption{The main panels show the wall clock speedup relative to serial execution for a model of a protoplanetary disk. The right panel shows the same model as that in the left panel, but with 10 times more photon packets for every iteration. The open circles show the average values of 10 executions of the code, and the uncertainties in these values derived from the scatter are not shown as they are smaller than the data points. The dashed curves show the speedup expected from a perfectly parallelized code. The solid curves show the best fit to the speedup curve assuming Amdahl's law (described in text). The gray shaded areas around the curves show the 68.3\% confidence interval. The maximum speedups derived from the best fits are shown as horizontal dotted lines. The inset panels show the probability of different $P$ values, with the 68.3\% confidence intervals shown in shaded gray.\label{fig:parallel_performance}}
\end{figure*}

\subsection{Additional user options}

\subsubsection{Uncertainties}

When computing SEDs and images, \hyperion allows the user to compute uncertainties, which uses the scatter in the photon packet flux values to derive errors in the flux at each wavelength/frequency and in each aperture or pixel. The fact that a given SED or image can be constructed using a combination of techniques, such as peeling-off and raytracing -- which can produce very different signal-to-noise -- is properly taken into account by computing the uncertainties for the contribution from each technique to the final SEDs or images separately and then combining them.

\subsubsection{Photon tracking}

\hyperion offers the option for the user to track the origin of each photon packet to split SEDs and images into different components. A basic mode allows SEDs and images to be split into contributions from sources and from dust, while a more detailed mode allows the flux to be split into individual sources and dust types. In both cases, the flux can be split further into photon packets reaching the observer directly, and photon packets having been scattered since the last emission/re-emission and before reaching the observer.

\subsubsection{Dust sublimation}

Users have the option to specify a dust sublimation specific energy absorption rate for each dust type, and three different dust sublimation modes are possible:

\begin{itemize}
\item the specific energy absorption rate can simply be capped to the maximum value, without changing the density.
\item the dust can be completely removed from cells exceeding the maximum specific energy absorption rate.
\item the dust density can be reduced but not set to zero. This can be useful because in optically thick cells, if radiation originates from the cell (for example luminosity from viscous dissipation), the specific energy absorption rate can be high because the radiation is trapped. If the density had been lower but non-zero, the specific energy absorption rate might not have exceeded the maximum specified. In this situation, the dust density should be reduced but not set to zero.
\end{itemize}

\subsubsection{Forced first scattering}

As mentioned in Section \ref{sec:introduction}, forced first scattering \citep[e.g.][]{mattila:70:53, wood:99:799} can be used to improve the signal-to-noise of scattered radiation in optically thin dust. \hyperion includes an implementation of this algorithm.

\section{Parallelized performance}

\label{sec:parallel}

The reason for not using the \citet{bjorkman:01:615} immediate temperature correction method in \hyperion is that each time a photon gets absorbed and re-emitted, the specific energy absorption rate grid has to be updated. Thus, it is not possible to compute the propagation of multiple photons simultaneously and to then combine the results, and codes using this method can therefore not easily be parallelized. By simply using an iterative approach to computing the specific energy absorption rate using the \citet{Lucy:99:282} continuous absorption method, one has the advantage that within an iteration, the problem is ``embarrassingly parallel''. Each process can propagate photon packets independently, and at the end of the iteration, the energy absorbed in each cell is synchronized over processes. Similarly, when SEDs, images, or polarization maps are computed, these can be computed by separate processes, and combined at the end of the calculation.

The code has been parallelized using the Message Passing Interface (MPI). Since different cores or nodes can have inhomogeneous performance, rather than dividing the total number of photon packets equally between processes, the task is divided into chunks of photon packets that are small enough that there are many more chunks than total number of processes, but large enough that the processes do not communicate too often. Each process then computes one batch of photon packets, and reports back to the main process to find out whether to process another batch or whether to stop. This incurs a very small overhead, since the request consists essentially of a single number both ways. Only at the end of the iteration, once all processes have received the signal to stop, are the results combined. This last step incurs an overhead, which scales with the number of processes and the grid, image, or SED resolution. However, in most cases, the overhead is negligible compared to the main computation.

The parallel efficiency of the code depends on the particular model being computed and the number of photon packets requested. One way to quantify the speedup for a given model is to compare the fraction of time spent in the serial execution of the code, such as the startup phase where the data is read in from the input file, or the time between iterations, to the fraction of time spent in the ``embarrassingly parallel'' photon packet propagation. If one writes the fraction of the code that is truly parallel as $P$, then the speedup obtained by running the code on $N$ parallel processes is:
\begin{equation}
\label{eq:amdahl}
{\rm speedup} = \frac{1}{(1-P) + P/N}
\end{equation}
This is often referred to as Amdahl's law \citep{Amdahl:67:483}.
One consequence of Equation (\ref{eq:amdahl}) is that as $N$ tends to infinity, the speedup tends to a finite maximum
\begin{equation}
{\rm speedup}_{\rm max} = \frac{1}{1-P}
\end{equation}
In this case, the parallelized part of the code runs infinitely fast, and the execution time is given purely by the serial portion of the code.

\begin{figure*}
\centering
\includegraphics[width=\textwidth]{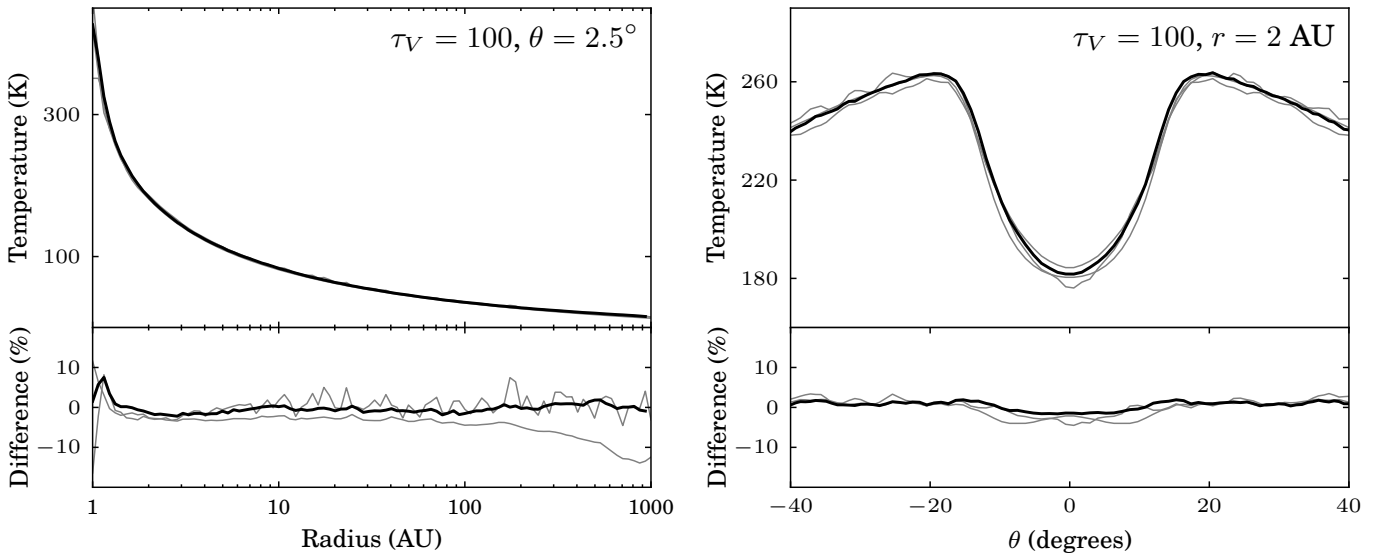}
\caption{Temperature results for the \cite{Pascucci:04:793} disk benchmark, for the $\tau_V=100$ model. The left panel shows a radial temperature profile close to the mid-plane ($\theta=2.5^\circ$), both in absolute terms (top) and relative to the reference code RADICAL (bottom). The right panel shows a vertical cut through the disk at a cylindrical radius of 2\,AU. The black lines show the results from \hyperion, while the gray lines show the results from the codes tested in \cite{Pascucci:04:793}}
\label{fig:pasc_temp}
\end{figure*}

\begin{figure*}
\center
\includegraphics[width=\textwidth]{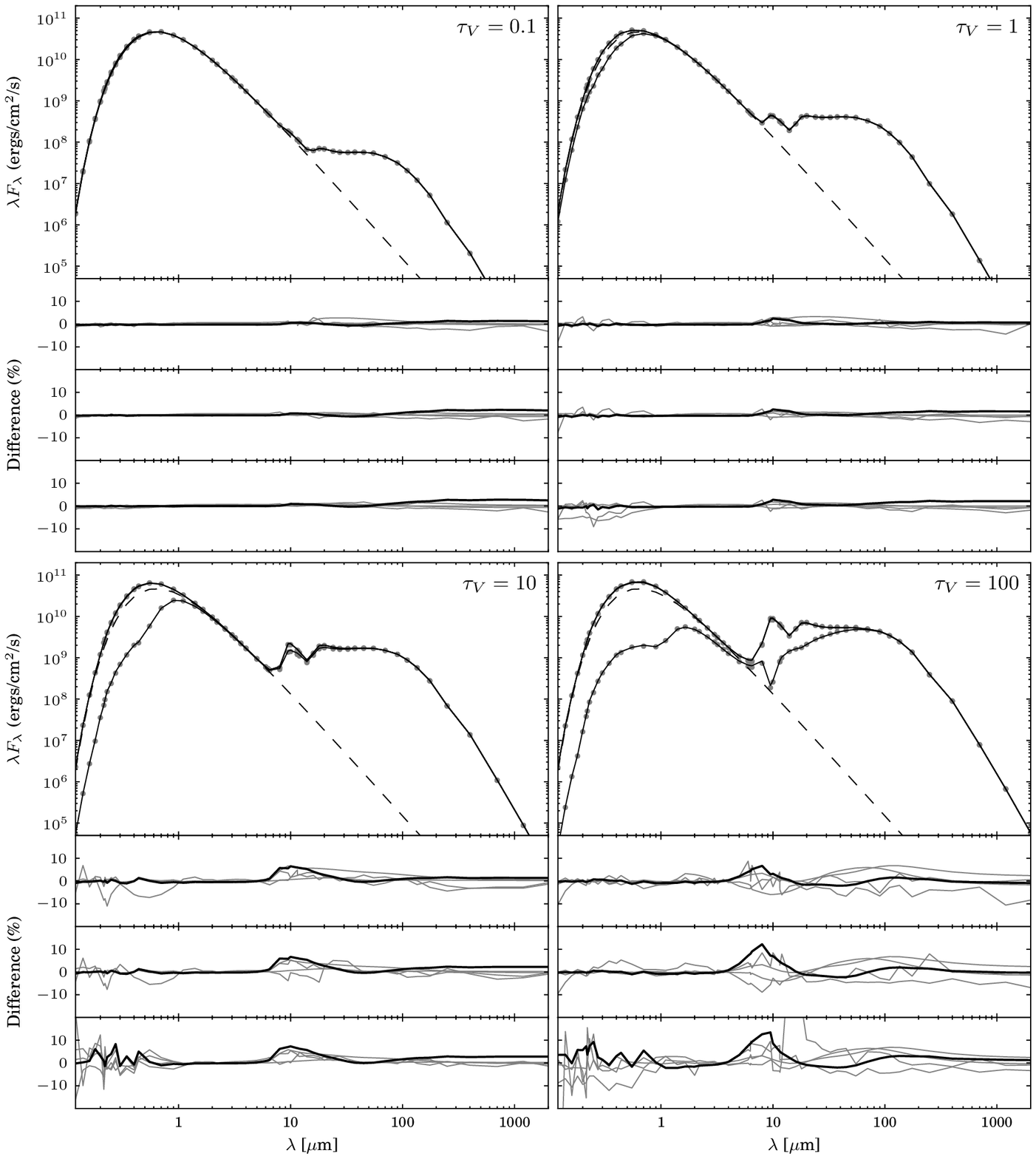}
\caption{SED results for the \cite{Pascucci:04:793} disk benchmark, for the four disk masses used. In each panel, the top section shows the SED obtained using \hyperion (black line), compared to the reference code from \cite{Pascucci:04:793} (RADICAL; gray circles), for the three viewing angles used (12.5, 42.5, and 77.5$^\circ$). The dashed line shows the blackbody spectrum of the central source. The bottom three sections in each panel show the fractional difference between \hyperion and RADICAL (black line), and between the other codes in \cite{Pascucci:04:793} and RADICAL (gray lines).\label{fig:pasc_seds}}
\end{figure*}

\begin{figure*}
\begin{center}
\includegraphics{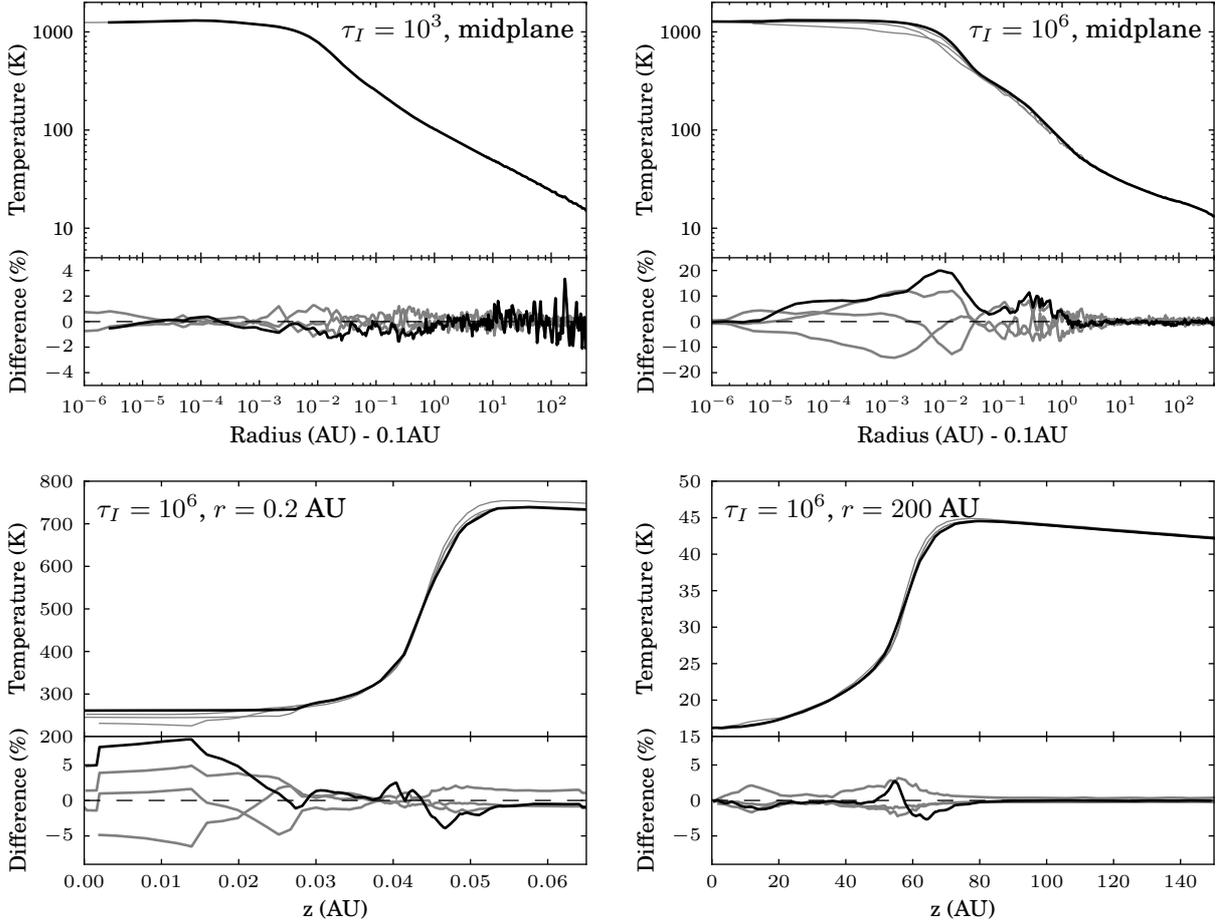}
\caption{Temperature results for the \cite{Pinte:09:967} disk benchmark. The top-left and top-right panels shows the radial temperature profile at the disk mid-plane for the lowest and highest disk mass models. The bottom panels shows two vertical cuts through the disk for the highest disk mass model, at cylindrical radii of 0.2 and 200\,AU respectively. In each panel, the top section shows the temperature profile in absolute terms, while the bottom section shows the results relative to the reference result from \cite{Pinte:09:967}, which is the average of the results from MCFOST, MCMAX, and TORUS. The black lines show the results from \hyperion, while the gray lines show the results from MCFOST, MCMAX, and TORUS.\label{fig:pinte_temp}}
\end{center}
\end{figure*}

\begin{figure*}
\begin{center}
\includegraphics{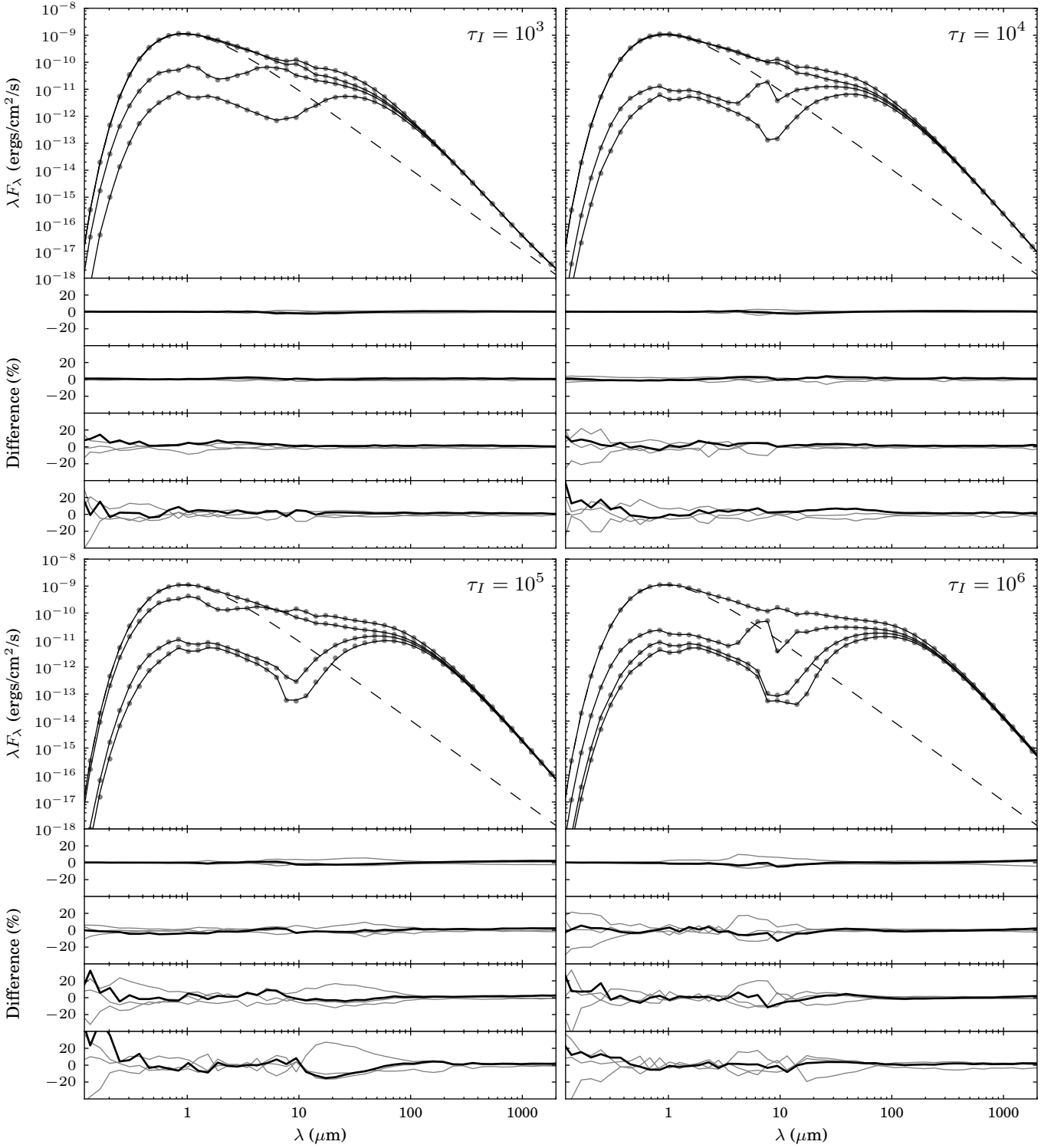}
\caption{SED results for the \cite{Pinte:09:967} disk benchmark, for the four disk masses used. In each panel, the top section shows the SED obtained using \hyperion (black line), compared to the average result of the codes used in \cite{Pinte:09:967} (gray circles), for four of the viewing angles (18.2, 75.5, 81.4, 87.1$^\circ$). The dashed line shows the blackbody spectrum of the central source. The bottom four sections in each panel show the fractional difference between \hyperion and the reference result (black line), and between the other codes compared in \cite{Pinte:09:967} and the reference result (gray lines).\label{fig:pinte_seds}}
\end{center}
\end{figure*}

\begin{figure*}
\begin{center}
\includegraphics{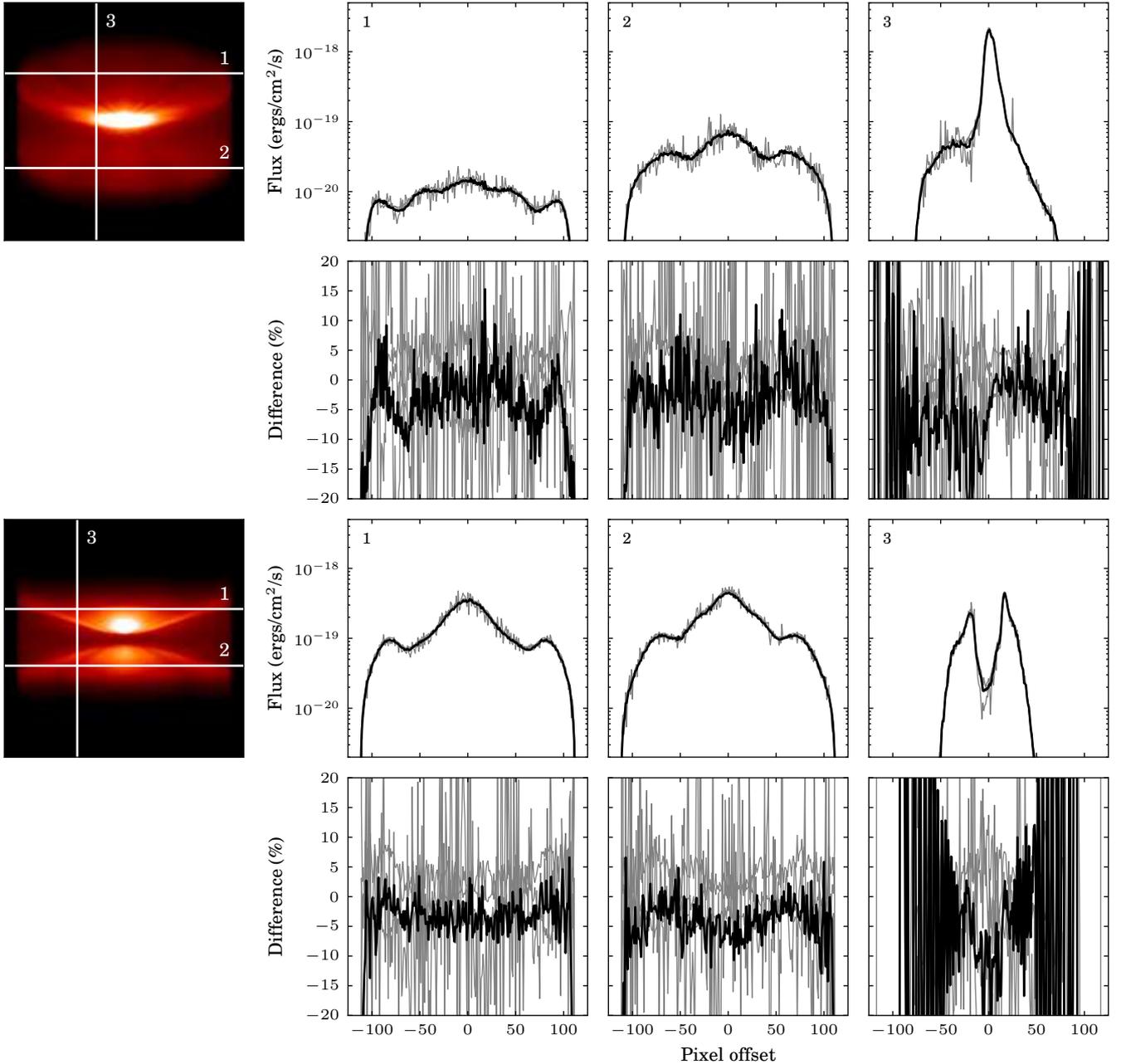}
\caption{Image results for the \cite{Pinte:09:967} disk benchmark. The top two rows show the results for the $i=69.5^\circ$ model, while the bottom two rows show the results for the $i=87.1^\circ$ model. For each viewing angle, the resulting image is shown on the left on a power-law stretch (with power 1/4), while the remaining plots show various cuts (indicated on the images), in absolute flux units, as well as relative to the reference result in \cite{Pinte:09:967}, which is the average of the MCFOST, MCMAX, TORUS, and Pinball codes. The black lines show the results from \hyperion, while the gray lines show the results from MCFOST, MCMAX, TORUS, and Pinball. As in \cite{Pinte:09:967}, the cuts are 11 pixels wide to improve the signal-to-noise of the profiles.\label{fig:pinte_I}}
\end{center}
\end{figure*}

\begin{figure*}
\begin{center}
\includegraphics{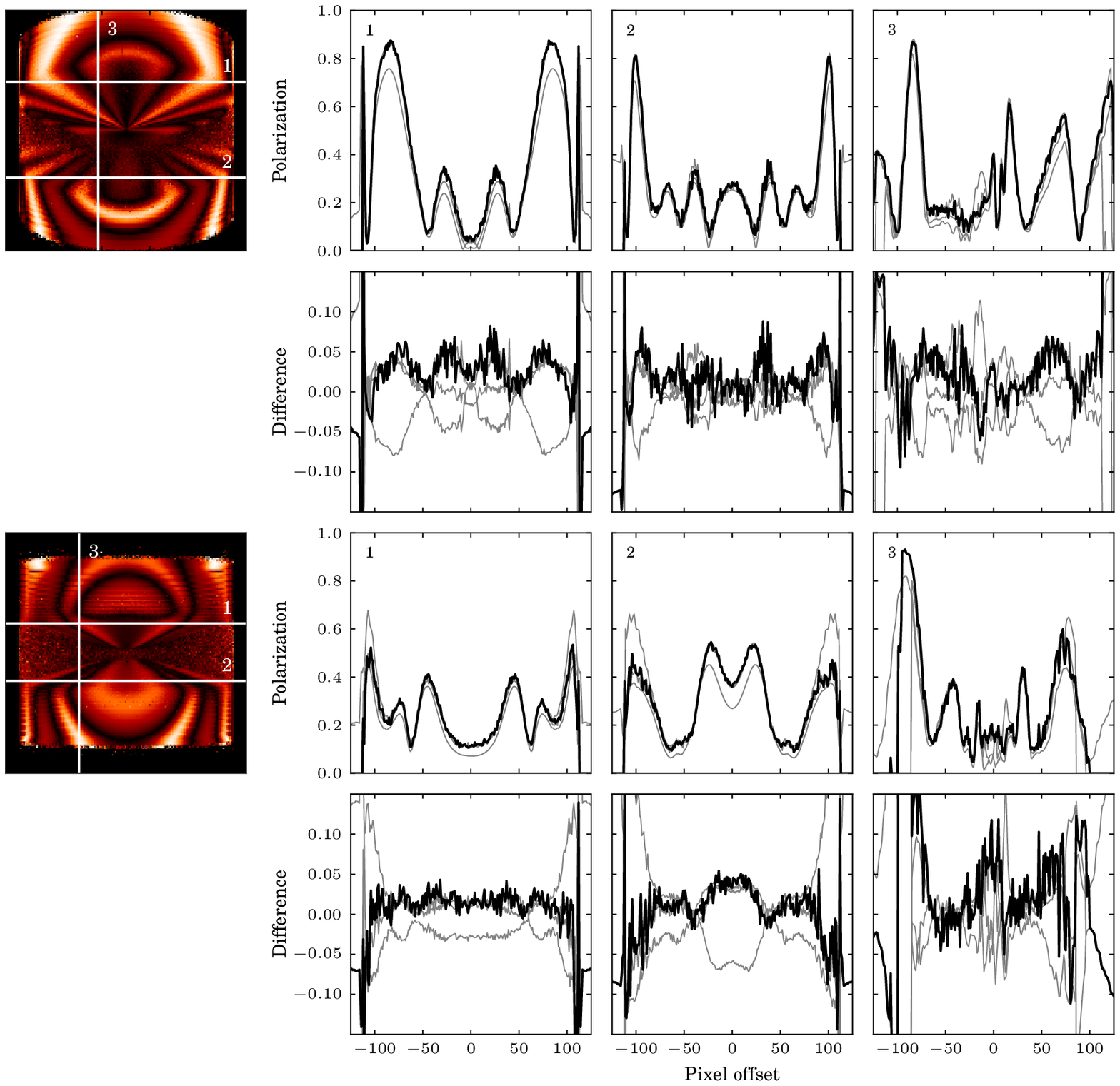}
\caption{Linear polarization map results for the \cite{Pinte:09:967} disk benchmark. The top two rows show the results for the $i=69.5^\circ$ model, while the bottom two rows show the results for the $i=87.1^\circ$ model. For each viewing angle, the resulting linear polarization map is shown on the left on a linear stretch, while the remaining plots show various cuts (indicated on the maps, and identical to those in Figure \ref{fig:pinte_I}), both in absolute terms, and relative to the reference result in \cite{Pinte:09:967}, which is the average of the MCFOST, MCMAX, and Pinball codes. The black lines show the results from \hyperion, while the gray lines show the results from MCFOST, MCMAX, and Pinball. As in \cite{Pinte:09:967}, the cuts are 11 pixels wide to improve the signal-to-noise of the profiles.\label{fig:pinte_pol}}
\end{center}
\end{figure*}

The value of $P$ is dependent on the choice of the model being computed and number of photon packets. This is demonstrated here  with a model of a protoplanetary disk (taken from \S\ref{sec:pinte}) computed on different numbers of cores, ranging from $N=1$ to $N=512$ in powers of 2. The wall clock times of the parallel computations are compared to the true serial runtime rather than the parallel version running with one process, since these will have slightly different runtimes. The times used are wall clock times\footnote{This can also be referred to as `real' time}, rather than CPU times. For each number of cores, the model was run ten times to obtain a mean speedup and a measure of the scatter from one run to the next. The results are shown in Figure \ref{fig:parallel_performance} for two instances of the model - one with ten times more photon packets than the other. The models were run on the Harvard Odyssey cluster, allowing N up to 512. For the model with the lower number of photon packets\footnote{For the model with the lower number of photon packets, the results for $N=512$ were not used, since the time to start up the processes on the nodes dominated the wall clock time, and therefore the speedup values obtained were a measure of the efficiency of the cluster than of the code}, the speedups obtained are well fit by $P=0.99658\pm0.00014$, while the speedups for the model with the higher number of photon packets were well fit by $P=0.999373\pm0.00006$ where the uncertainties were derived as a formal confidence interval using the $\chi^2$ surface. The reduced $\chi^2$ values for the fits were 1.01 and 3.25 respectively, indicating good fits. The $P$ values translate into maximum speedups of $292.6^{+12.3}_{-11.5}$ and $1604.4^{+174.8}_{-147.3}$ respectively. These values differ by almost an order of magnitude, which is expected, since the only difference between the two models is that the CPU time of the parallel section of the code was increased by a factor of ten, while the serial part of the code remained identical (same input/output and synchronization overheads). In other words, both models tend to the same runtime, since the serial portion of the code is the same in both cases.

The speedup values presented here are purely illustrative, and should not be assumed to hold for all models, but they serve to show that the efficiency of the code does have a finite limit which depends on the model set-up. As for most parallel codes, beyond a certain number of parallel processes, the wall clock runtime is completely dominated by the time to execute the serial part of the code (typically input/output and calculations between iterations such as the PDA). Users are encouraged to determine the optimal number of processes for the problem they are considering.

\begin{figure*}
\begin{center}
\includegraphics[width=\textwidth]{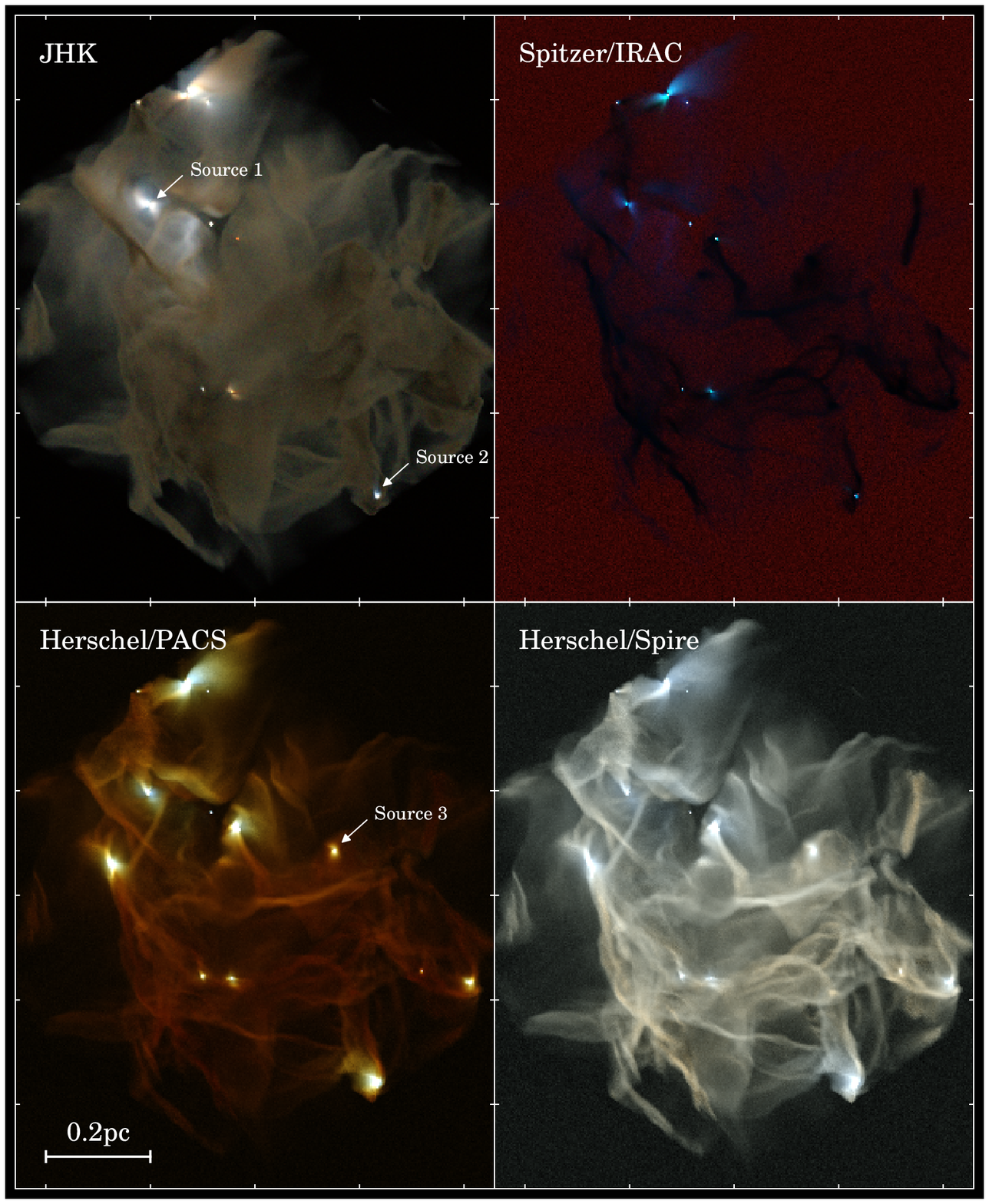}
\caption{Three-color unconvolved synthetic images of the simulation, from the near-infrared (top left) and mid-infrared (top right) to the far-infrared (bottom left and right). The wavelengths and colors used are 1.25\,$\mu$m (blue), 1.65\,$\mu$m (green), and 2.15\,$\mu$m (red) for the JHK image, 3.6\,$\mu$m (blue), 4.5\,$\mu$m (green), and 8.0\,$\mu$m (red) for the \textit{Spitzer}/IRAC image, 70\,$\mu$m (blue), 110\,$\mu$m (green), and 170\,$\mu$m (red) for the \textit{Herschel}/PACS image, and 250\,$\mu$m (blue), 350\,$\mu$m (green), and 500\,$\mu$m (red) for the \textit{Herschel}/SPIRE image. All images are shown on an arcsinh stretch which allows a larger dynamic range to be shown by reducing the intensity of the brightest regions. The exact transformation is given by $v_{\rm max}\cdot{\rm asinh}\left(m\,v/v_{\rm max}\right)/{\rm asinh}\left(m\right)\,$ where $v$ is the original flux, $m=30$ is a parameter controlling the compression of the dynamic range, and $v_{\rm max}$ is 6, 4, and 2\,MJy/sr for J, H, and K; 2, 2, and 4\,MJy/sr for 3.6\,$\mu$m, 4.5\,$\mu$m, and 8.0\,$\mu$m; 3000\,MJy/sr for the three PACS bands; and 1000, 500, and 250\,MJy/sr for SPIRE 250\,$\mu$m, 350\,$\mu$m, and 500\,$\mu$m respectively.
\label{fig:synthetic_uniform}}
\end{center}
\end{figure*}

\begin{figure*}
\begin{center}
\includegraphics[width=\textwidth]{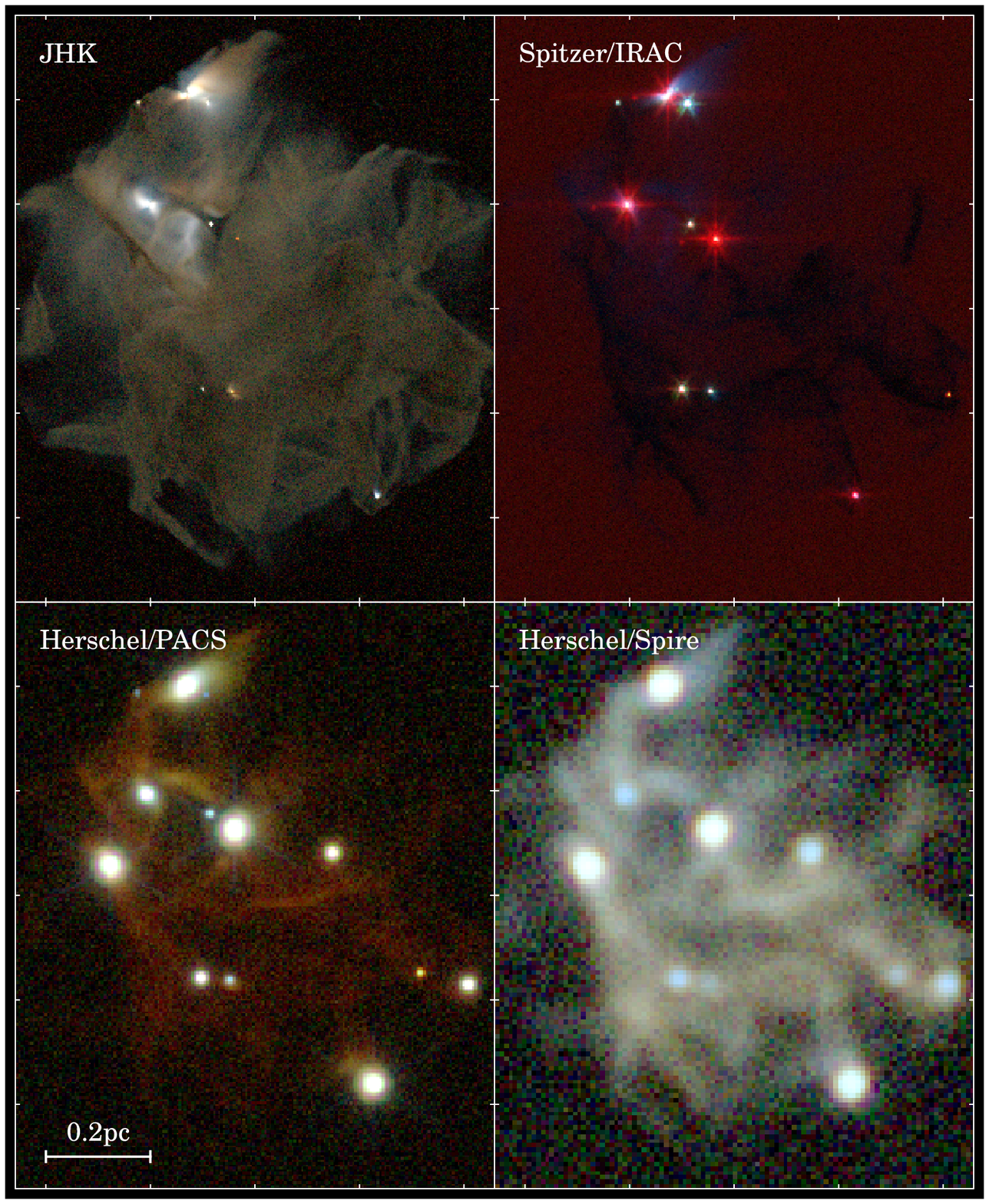}
\caption{Three-color synthetic images of the simulation, from the near-infrared (top left) and mid-infrared (top right) to the far-infrared (bottom left and right). The images are binned to the pixel resolutions appropriate for ground based observations (JHK) and for \textit{Spitzer} and \textit{Herschel} observations respectively. The images were convolved with the appropriate instrumental PSFs, and include Gaussian noise with realistic levels. The wavelengths, stretches, and levels used for the images are identical to those used in Figure \ref{fig:synthetic_uniform}.\label{fig:synthetic_convolved}}
\end{center}
\end{figure*}

\section{Benchmarks}

\label{sec:benchmarks}

\subsection{\cite{Pascucci:04:793} benchmark}

\subsubsection{Definition}

\cite{Pascucci:04:793} defined a benchmark problem for 2D radiative transfer, consisting of a flared disk around a point source. The central source is a point source with a 5,800\,K blackbody spectrum, and a luminosity of 1.016\,L$_\odot$. The disk extends from 1 to 1,000\,AU, and has a density structure given by:
\begin{equation}
\label{eq:disk}
\rho(r, z) = \rho_0\,\left(\frac{r}{r_0}\right)^{-\alpha}\,\exp{\left[-\frac{1}{2}\,\left(\frac{z}{h_0 \left(r/r_0\right)^{\beta}}\right)^2\right]},
\end{equation}
where r is the cylindrical polar radius, $r_0=500$\,AU, $h_0=125\cdot\sqrt{2/\pi}$\,AU, $\alpha=1.$, and $\beta=1.125$. The disk is made up of dust grains with a single size of 0.12\,$\mu$m, a density of $3.6$\,g/cm$^3$, and composed of astronomical silicates with properties given by \cite{Draine:84:89}. Scattering is assumed to be isotropic.

The benchmark consists of four cases, with visual ($\lambda=550$\,nm) optical depths, measured radially along the disk mid-plane, of $\tau_V=$0.1, 1, 10, and 100. These correspond to disk (dust) masses of $1.1138\times10^{-7}$, $1.1138\times10^{-6}$, $1.1138\times10^{-5}$, and $1.1138\times10^{-4}$\,M$_\odot$ respectively.
The corresponding maximum optical depths perpendicular to the disk are of the order of $\tau_V=$0.01, 0.1, 1, and 10 respectively.
SEDs are computed at three viewing angles in each case: 12.5, 42.5, and 77.5$^\circ$.

\subsubsection{Results}

The benchmark models were computed using a spherical polar grid with dimensions $(n_r, n_\theta, n_\phi)=(499, 399, 1)$ and extending out to 1,300\,AU. The SEDs were computed at specific wavelengths using the monochromatic radiative transfer described in \S\ref{sec:seds}. The MieX code \citep{Wolf:04:113} was used to compute the optical properties of the dust. The convergence algorithm discussed in \S\ref{sec:convergence} was used, with $p=99$\%, $Q_{\rm thres}=2$, and $\Delta_{\rm thres}=1.02$. The models were run with enough photon packets to provide very high signal-to-noise in the temperatures and SEDs\footnote{$10^7$ photons for each specific energy iteration, $10^5$ photons per wavelength for the monochromatic peeling-off for scattering for both the source and dust emission, and $10^6$ photons for the source and dust emission raytracing.}, and were run using the parallel version of the code on a single machine with 8 cores, requiring a wall clock time of under 20 minutes for each model. The $\tau_V=$0.1 and 1 models converged after $3$ iterations, while the $\tau_V=$10 and 100 models converged after $4$ iterations.

Figure \ref{fig:pasc_temp} shows for the most optically thick case ($\tau_V=100$) the temperature profile for a fixed polar angle ($\theta=2.5^\circ$) as a function of radius $r$, and the temperature profile for a fixed radius ($r=2$\,AU) as a function of polar angle $\theta$. The temperatures found by \hyperion are within the dispersion of the results from the other codes.

Figures \ref{fig:pasc_seds} shows the SEDs for the \hyperion code and the reference code in \citeauthor{Pascucci:04:793} (RADICAL) for the four disk masses and three viewing angles, and the fractional difference between the SEDs is shown in each case. Also shown in gray are the differences for other codes presented in \citeauthor{Pascucci:04:793}. The SEDs from \hyperion are within the dispersion of results from the other codes.

\subsection{\cite{Pinte:09:967} benchmark}

\label{sec:pinte}

\subsubsection{Definition}

While the \cite{Pascucci:04:793} benchmark includes a case with fairly high visual optical depth through the mid-plane of the disk, the optical depths through realistic protoplanetary disks are much higher ($\tau_V > 10^6$ in some cases), and are such that computing temperatures in the disk mid-plane becomes computationally challenging and requires approximations to be made (e.g. \S\ref{sec:propagation} and \S\ref{sec:temperature}). \cite{Pinte:09:967} developed a complementary benchmark problem consisting of more massive disks with a more extreme radial density profile to test codes in the limit of very optically thick disks. In addition, the benchmark also tests the ability to compute anisotropic scattering, and to reproduce images and polarization maps at various viewing angles.

The benchmark case consists once again of a central star surrounded by a disk. The central star has a radius of 2\,R$_\odot$ and a temperature of 4000\,K, with a blackbody spectrum. The disk extends from 0.1 to 400\,AU (with cylindrical edges), and the density is given by Equation (\ref{eq:disk}), with $r_0=100$\,AU, $h_0=10$\,AU, $\alpha=2.625$, and $\beta=1.125$. The disk is made up of dust grains with a single size of 1\,$\mu$m, a density of $3.5$\,g/cm$^3$, and composed of astronomical silicates with properties given by \cite{Weingartner:01:296}.

The benchmark consists of four cases, with disk (dust) masses of $3\times10^{-8}$, $3\times10^{-7}$, $3\times10^{-6}$, and $3\times10^{-5}$\,M$_\odot$. While the disk masses are comparable to that of the Pascucci et al. benchmark problem, the smaller inner radius (0.1 vs 1\,AU) and the much steeper surface density profile make this a more computationally challenging problem. In fact, the I-band ($\lambda=810$\,nm) optical depths, measured radially along the disk mid-plane, are of the order of $\tau_I=10^3$, $10^4$, $10^5$, and  $10^6$ for the four disk masses respectively, orders of magnitude higher than for the Pascucci et al. benchmark.
The corresponding maximum optical depths perpendicular to the disk are of the order of $\tau_I=100$, $10^3$, $10^4$, and $10^5$ respectively.

The benchmark test compares SEDs, images and polarization maps at 10 viewing angles corresponding to $\cos{i}=0.05$ to $0.95$ in steps of $0.10$, corresponding to viewing angles from $18.2$ (almost pole-on) to $87.1^\circ$ (almost edge-on). The images and polarization maps are computed at 1\,$\mu$m, and span $251\times251$ pixels, and a physical size of $900\times900$\,AU. Due to the single grain size, the scattering phase function oscillates strongly around 1\,$\mu$m and at shorter wavelengths. Thus, these oscillations have a direct impact on scattered-light images computed at these wavelengths. While a single grain size, and therefore these oscillations, would not be seen in a real disk, this allows codes to test whether they correctly compute the scattering of photon packets.

\subsubsection{Results}

The benchmark models were computed using a cylindrical polar grid with dimensions $(n_r, n_z, n_\phi)=(499, 399, 1)$ extending out to the outer disk radius. The SEDs were computed at exact wavelengths using the monochromatic radiative transfer described in \S\ref{sec:seds}. As before, the MieX code was used to compute the optical properties of the dust. The convergence algorithm discussed in \S\ref{sec:convergence} was used, with $p=99$\%, $Q_{\rm thres}=2$, and $\Delta_{\rm thres}=1.02$. The MRW (with $\gamma=2$) and PDA approximations were used due to the high optical depths. The models were run with enough photon packets to provide very high signal-to-noise in the temperatures, SEDs, and images\footnote{$10^8$ photons for each specific energy iteration, $10^7$ and $2\times10^8$ photons per wavelength for the monochromatic peeling-off for scattering of source and dust emission respectively (for SEDs; 100 times more for the images and polarization maps), and $10^7$ photons for the raytracing for both the source and dust emission.}, and were run using the parallel version of the code on 32 cores, requiring a wall clock time of 12 to 17 hours to compute the temperature and SEDs for the $\tau_I=10^3$ to $10^6$ models, and an additional 11 hours for the images and polarization maps. The temperatures converged after 7, 9, 9, and 10 iterations for the $\tau_I=10^3$, $10^4$, $10^5$, and  $10^6$ models respectively.

Figure \ref{fig:pinte_temp} compares the mid-plane temperature profile in two of the models ($\tau_I=10^3$ and $\tau_I=10^6$), as well as two vertical cuts in the most optically thick model ($\tau_I=10^6$). Figure \ref{fig:pinte_seds} compares the SEDs for the four models at four of the viewing angles (18.2, 75.5, 81.4, 87.1$^\circ$). Finally, Figures \ref{fig:pinte_I} and \ref{fig:pinte_pol} compare the total intensity and polarization fraction maps for the most optically thick model ($\tau_I=10^6$). In all cases, the results produced by \hyperion are within the dispersion of results obtained from the other codes.

\section{Case study: simulated observations of a low-mass star-forming region}

\label{sec:simulation}

\begin{table}
\caption{Noise levels added to the synthetic images in Figure \ref{fig:synthetic_convolved}. \label{tab:synthetic}}
\centering
\begin{tabular}{cc}
\hline\hline
Band & Noise $\sigma$ \\
     & (MJy/sr) \\
\hline
J & 0.03 \\
H & 0.03 \\
K & 0.03 \\
IRAC 3.6\,$\mu$m & 0.005 \\
IRAC 4.5\,$\mu$m & 0.005 \\
IRAC 8.0\,$\mu$m & 0.03 \\
PACS 70\,$\mu$m & 20. \\
PACS 110\,$\mu$m & 30. \\
PACS 170\,$\mu$m & 40. \\
SPIRE 250\,$\mu$m & 15. \\
SPIRE 350\,$\mu$m & 10. \\
SPIRE 500\,$\mu$m &  5. \\
\hline
\end{tabular}
\end{table}

\subsection{The simulation}

To demonstrate the capabilities of the code, a radiative transfer calculation on a simulation of a low-mass star forming cloud was carried out in order to produce synthetic observations from near-infrared to sub-milimeter wavelengths.

The demonstration uses a simulation of low-mass star formation computed with the ORION AMR three-dimensional gravito-radiation-hydrodyamics code \citep[][and references therein]{offner:09:131}. The simulation contains 185\,M$_\odot$ of gas and dust in a box of side 0.65\,pc, and resolves size-scales down to 32\,AU, with an effective resolution of $4096^3$. \citeauthor{offner:09:131} studied the effects of heating from protostars on the formation of stars, and therefore ran a simulation including radiative heating, and a control case with no radiation heating. The simulation used here is a snapshot of the one including radiative heating, for $t\sim t_{\rm ff}$, where $t_{\rm ff}$ is the free-fall time.

\begin{figure*}
\begin{center}
\includegraphics[width=\textwidth]{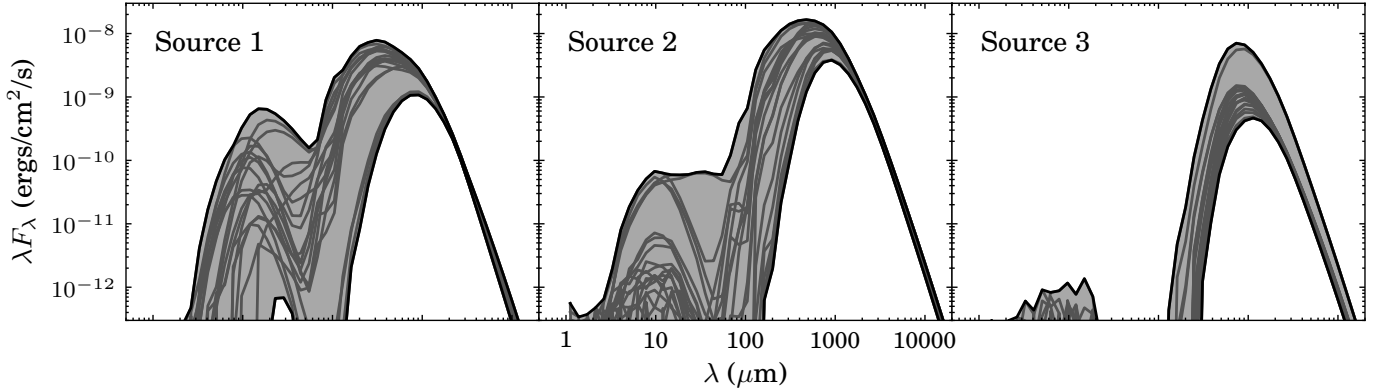}
\caption{Spectral Energy Distributions for the three sources labelled in Figure~\ref{fig:synthetic_uniform}. Each panel shows SEDs for 20 different viewing angles. The black lines show the minimum and maximum values as a function of wavelength, the light gray shows the range of values, and the dark gray shows the individual SEDs. \label{fig:synthetic_seds}}
\end{center}
\end{figure*}

\subsection{The radiative transfer model}

The radiative transfer through the star-forming region was simulated by taking into account both the luminosity from the forming stars (c.f. Appendix B of \citealt{offner:09:131}), represented by sink particles in the simulation, and the interstellar radiation field at the solar neighborhood from \citet{Porter:05:77}, which includes contributions from the stellar emission, PAH emission, and far-infrared thermal emission\footnote{The stellar component provides photon packets that are scattered off the cloud, but the direct (un-scattered) emission from this component is not included in the images presented in Figures \ref{fig:synthetic_uniform} and \ref{fig:synthetic_convolved} since it would not appear as a diffuse component but as point sources.}. The simulation was embedded at the center of a cube of side 1.95\,pc and a uniform density of $10^{-22}$ g/cm to simulate an ambient medium. The dust properties were taken from \citet{Draine:03:241, Draine:03:1017} using the \citet{Weingartner:01:296} Milky Way grain size distribution A for $R_{\rm V}$=5.5 and $C/H=30$\,ppm renormalized to $C/H=42.6$\,ppm. In order to compute the full Mie scattering properties of this dust model, the \texttt{bhmie} routine from \citep{Bohren:83} and modified by B. Draine\footnote{\url{http://www.astro.princeton.edu/~draine/scattering.html}} was used.

The radiative equilibrium temperatures were first computed, followed by images at 12 wavelengths ranging from 1.25 microns to 500 microns\footnote{The calculation used $10^8$ photons for each specific energy iteration, $10^9$ photons per wavelength for the monochromatic peeling-off for scattering for both the source and dust emission, and $10^9$ photons for the source and dust emission raytracing.}. The wavelengths were chosen to produce synthetic images in common bands, including JHK (1.25, 1.65, and 2.15\,$\mu$m), \textit{Spitzer}/IRAC 3.6, 4.5, and 8.0\,$\mu$m, \textit{Herschel}/PACS (70, 110, and 170\,$\mu$m), and \textit{Herschel}/SPIRE (250, 350, and 500\,$\mu$m). A distance of 300\,pc was used to simulate a nearby low-mass star-formation region.
The images were computed at a viewing angle of $(\theta, \phi)=(45^\circ, 45^\circ)$, and are shown in Figure \ref{fig:synthetic_uniform} at a common resolution of 1\arcsec/pix.

The near-infrared (JHK) image shows emission from the diffuse cloud, which is scattered light from the interstellar radiation field, as well as bright features corresponding to light from the forming stars scattered in the circumstellar dust. The near-infrared images are entirely dominated by direct and scattered stellar light, and there is no contribution from thermal emission. The mean colors of the near-infrared emission are J-H$=0.387\pm0.280$ and H-K$=-0.214\pm0.273$ where the uncertainties indicate the scatter in the values rather than the error in the mean.

The IRAC image also shows (in blue) the scattered light from some of the sources, and a couple of the sources show a bipolar structure. This is noteworthy because the simulation does \textit{not} include outflows. The bipolar structures seen here are instead due to beaming of the radiation by the three-dimensional structure of the accretion flows and the presence of protostellar disks. These cause the density to be lower along the rotation axis, and the radiation to preferentially escape in those directions. Thus, bipolar scattered light structures in the near- and mid-infrared are not necessarily evidence for actual outflows.

The IRAC image also shows the bright PAH background from the interstellar radiation field, which causes the cloud to appear in absorption as an infrared-dark cloud (IRDC). The column density of the filaments is typical of IRDCs: for instance, 10\% of the area shown in Figure \ref{fig:synthetic_uniform} has $\Sigma>0.10$\,g/cm$^2$ or $A_{\rm V}>34.6$\,mag and 1\% of the area has $\Sigma>0.27$\,g/cm$^2$ or $A_{\rm V}>90.7$\,mag. In contrast, the far-infrared images show the cloud in emission. The color gradients across the PACS image trace gradients in the dust temperature, with the regions immediately surrounding the protostars being warmer than the rest of the cloud. Finally, the SPIRE image traces most of the mass, but does not show very strong color gradients, as most of the emission is in the Rayleigh-Jeans part of the spectrum.

\begin{table*}
\caption{Properties of the sources in Figure \ref{fig:synthetic_seds}.\label{tab:sed_sources}}
\centering
\begin{tabular}{cccccccc}
\hline\hline
Source & $M_\star$ & $\dot{M}$ & $\dot{M}/M_\star$ & $L_\star$ & $L_{\rm acc}$ & $L_{\rm disk}$ & $L_{\rm tot}$ \\
 & ($M_\odot$) & ($M_\odot\cdot$yr$^{-1}$) & (yr$^{-1}$) & $L_\odot$ & $L_\odot$ & $L_\odot$ & $L_\odot$ \\
\hline
1  & 2.216 & 6.012e-07 & 2.713e-07 & 24.20 & 1.25 &  2.51 & 27.96 \\
2a & 0.432 & 3.192e-06 & 7.388e-06 &  0.03 & 2.81 &  5.62 &  8.46 \\
2b & 1.548 & 4.089e-06 & 2.642e-06 &  5.32 & 9.74 & 19.48 & 34.54 \\
3  & 0.058 & 4.771e-06 & 8.163e-05 &  0.00 & 1.45 &  2.91 &  4.36 \\
\hline
\end{tabular}
\end{table*}

\subsection{Synthetic Observations}

Synthetic observations were produced  by resampling the pixel resolution of the images to values typical of ground based observations for JHK (1\arcsec), and of \textit{Spitzer} and \textit{Herschel} observations for the other bands (1.2\arcsec\,for IRAC, 3.2\arcsec\,for PACS, and 6\arcsec\,for SPIRE). The observations were convolved with the appropriate instrumental PSFs and Gaussian noise with typical values (listed in Table~\ref{tab:synthetic}) was added to the images. The resulting synthetic observations are shown in Figure \ref{fig:synthetic_convolved}.

The near-infrared (JHK) image does not change, since the pixel resolution is the same as before, and the PSF full-width at half maximum (1\arcsec) is the same as the pixel resolution. The IRAC image shows many of the protostars appearing as strong 8\,$\mu$m point sources. These were not seen in Figure \ref{fig:synthetic_uniform} since only one pixel contained this bright emission for each source. The PACS and SPIRE images are severely degraded, with only the brightest diffuse emission seen in the PACS image. Nevertheless, the PACS image shows very strong compact emission associated with the protostars. The SPIRE image does show bright point sources for a few of the protostars, but the remainder are confused with the cloud.

\subsection{Spectral Energy Distributions}

For three sources, indicated in Figure \ref{fig:synthetic_uniform}, SEDs were computed inside 20 apertures, with a central circular aperture 5,000\,AU in radius, and the remaining apertures consisting of annuli logarithmically spaced between 5,000\,AU to 20,000\,AU. The SEDs were directly computed by binning photon packets into apertures in the radiative transfer code (rather than using the images). In reality, measuring SEDs from convolved and noisy images is going to affect the resulting SEDs, but since this is highly dependent on the convolution and noise parameters, we do not take this into account here as we are interested in the `intrinsic' SEDs. The SEDs were computed for 20 viewing angles regularly distributed in three dimensions around the sources\footnote{Truly regular non-random three-dimensional spacing of points on a sphere is only possible for 4, 6, 8, 12, and 20 points, and the position of the points corresponds to the vertices of regular polyhedra. In the current section, the regular spacing is achieved by using the 20 vertices of a dodecahedron.}. The innermost circular aperture was used to compute the SED, and the median of the SEDs in the 19 surrounding annular apertures was used as an estimate of the wavelength-dependent background that was subtracted from the central aperture SED.

The resulting SEDs are shown in Figure \ref{fig:synthetic_seds}. The three sources were chosen as they were expected to produce different SEDs based on the images shown in Figure \ref{fig:synthetic_uniform}, and indeed the three sources show SEDs with varying amounts of near- and mid-infrared emission. The dependence of the near- and mid-infrared emission on viewing angle is very large, and is a consequence of the three-dimensional structure of the circumstellar material. The basic evolutionary properties of the sources are listed in Table~\ref{tab:sed_sources}. Source 2 is actually composed of a pair of sink particles, so these are listed as 2a and 2b in the table. Sources 1, 2a, and 2b have similar masses, of the order or $0.5-2$\,$M_\odot$. Source 3 in contrast has a much lower mass, around 0.06\,$M_\odot$. By looking at the accretion rate of the sources relative to the sink particle masses, one can see that the relative accretion rate increases from Source 1 to 3, which is consistent with the average behavior of the near- to far-infrared ratio in the SEDs. However, the evolutionary stage of the sources -- both relative to each other and in absolute terms -- may be derived incorrectly in reality, when only one viewing angle is available.

\section{Future}

\label{sec:future}

The code will be actively developed in the future to improve existing features and add new capabilities. Some of the planned capabilities include:

\begin{itemize}

\item Scattering and absorption by non-spherical grains aligned along magnetic fields. The main effect of aligned grains is to produce linear and polarization polarization signatures that trace the magnetic fields \citep{whitney:02:205}.

\item Continuum and line gas radiative transfer, including photoionization. This will be very useful for modeling ionization and dynamics in a wide range of environments.

\item Support for unstructured meshes, where each cell is an irregular polyhedron, and which are commonly constructed from a Voronoi tessellation of space. They allow a more precise sampling of arbitrary three-dimensional density structures than other types of adaptive grids for a fixed number of cells. Unstructured meshes are now being used in hydrodynamical simulations \citep[e.g.][]{springel:11:203, greif:11:75}, and support for these in \hyperion would allow radiative transfer post-processing of the simulations without the need for re-gridding the density structures.

\item Raytracing for scattered light, which - while memory intensive - would provide high signal-to-noise for both images and SEDs efficiently at wavelengths dominated by scattered light faster than the peeling-off algorithm \citep{Pinte:09:967}.

\item Temperature dependent opacities for dust, since the composition of dust often depends on the temperature, especially when including ices.

\end{itemize}

\section{Summary}

\hyperion is a new radiative transfer code that has been designed to be applicable to a wide range of problems. Radiative transfer can be carried out through a variety of three-dimensional grids, including cartesian, cylindrical and spherical polar, and adaptive cartesian grids. The algorithms used by the code, including the photon packet propagation, raytracing, convergence detection, and diffusion approximations are described in detail (\S\ref{sec:overview}).

The code is parallelized and scales well to hundreds or thousands of processes. As expected, the maximum speedup depends on the fraction of the execution time spent in the serial part of the code (such as file input/output), which in turn depends on the problem being computed (\S\ref{sec:parallel}).

The code was tested against the \citet{Pascucci:04:793} and \citet{Pinte:09:967} benchmarks for circumstellar disks and was found to be in excellent agreement with the respective published benchmark results (\S\ref{sec:benchmarks}).

As a case study, \hyperion was used to predict near-infrared to sub-millimeter observations of a star formation region using a dynamical simulation of a region of low-mass star formation \citep{offner:09:131}. One interesting finding was that despite the absence of outflows in the simulation, three-dimensional accretion flows caused beaming of radiation that produced fairly symmetric structures usually associated with bipolar outflow cavities.

\hyperion is published under an open-source license at \url{http://www.hyperion-rt.org}, using a version control hosting website that makes it easy for users to contribute to the code and documentation.

\section*{Acknowledgments}

I wish to thank the referee for a careful review and for comments that helped improve this paper.
I also wish to thank C. Pinte and I. Pascucci for assistance with their respective disk benchmark problems, Stella Offner for providing the star formation simulation and for useful discussions and comments,
and Barbara Whitney and Kenneth Wood for helpful discussions and comments.
The computations in this paper were run on the Odyssey cluster supported by the FAS Sciences Division Research Computing Group. The parallel performance tests were run using 0.7.7 of the code, while the benchmarks problems and the case study model were computed using version 0.8.4.
Support for this work was provided by NASA through the Spitzer Space Telescope Fellowship Program, through a contract issued by the Jet Propulsion Laboratory, California Institute of Technology under a contract with NASA.
This research has made use of NASA's Astrophysics Data System.
No photon packets were harmed in the making of this paper.

\bibliographystyle{apj_tr9}
\bibliography{apj-jour,ms_auto,extra}

\appendix

\onecolumn

\clearpage

\section{The partial diffusion approximation (PDA)}

\label{app:pda}

The diffusion approximation is described by
\begin{equation}
\nabla\cdot\left(D\nabla T^4\right) = 0,
\end{equation}
where $D=1/3\rho\bar{\chi_{\rm R}}$ is the diffusion coefficient. This equation can be written as a system of linear equations. Since \hyperion stores the specific energy absorption rate of the dust rather than the temperature, the PDA is actually solved using:

\begin{equation}
\nabla\cdot\left[D\nabla \left(\frac{\dot{A}}{\kappa_P(\dot{A})}\right)\right] = 0.
\end{equation}

In spherical polar coordinates, the diffusion equation is

\begin{equation}
\frac{1}{r^2}\frac{\partial}{\partial r}\left(r^2 D\,\frac{\partial T^4}{\partial r}\right)
+ \frac{1}{\sin{\theta}}\,\frac{\partial}{r\partial \theta}\left(\sin{\theta}\,D\,\frac{\partial T^4}{r\partial \theta}\right)
+ \frac{\partial}{r\sin{\theta}\partial \phi} \left(D\, \frac{\partial T^4}{r\sin{\theta}\partial \phi}\right )= 0.
\end{equation}
This equation can be written for a discrete spherical polar grid as
\begin{equation}
\begin{split}
\frac{1}{r^2_{i}\,\Delta r_{i}}
\left(
r^2_{i+1/2}\,\frac{T^4_{i+1,j,k} - T^4_{i,j,k}}{\Delta\tau^{r}_{i+1,j,k} + \Delta\tau^{r}_{i,j,k}}
+
r^2_{i-1/2}\,\frac{T^4_{i-1,j,k} - T^4_{i,j,k}}{\Delta\tau^{r}_{i,j,k} + \Delta\tau^{r}_{i-1,j,k}}
\right)\\
\frac{1}{\sin{\theta_{j}}}
\frac{1}{r\Delta \theta_{j}}
\left(
\sin{\theta}_{j+1/2}
\frac{T^4_{i,j+1,k} - T^4_{i,j,k}}{\Delta\tau^{\theta}_{i,j+1,k} + \Delta\tau^{\theta}_{i,j,k}}
+
\sin{\theta}_{j-1/2}
\frac{T^4_{i,j-1,k} - T^4_{i,j,k}}{\Delta\tau^{\theta}_{i,j,k} + \Delta\tau^{\theta}_{i,j-1,k}}
\right) & \\
\frac{1}{r\sin{\theta}\Delta \phi_{k}}
\left(
\frac{T^4_{i,j,k+1} - T^4_{i,j,k}}{\Delta\tau^{\phi}_{i,j,k+1} + \Delta\tau^{\phi}_{i,j,k}}
+
\frac{T^4_{i,j,k-1} - T^4_{i,j,k}}{\Delta\tau^{\phi}_{i,j,k} + \Delta\tau^{\phi}_{i,j,k-1}}
\right)
& = 0,
\end{split}
\end{equation}
where $i$, $j$, and $k$ are the indices of the grid cells, $r$, $\theta$ and $\phi$ are the coordinates of the center of the cell, $\Delta r_i$, $r\Delta \theta_j$, and $r\sin{\theta}\Delta \phi_k$ are the spatial extent of the cell along each coordinate, and the $\Delta\tau$ terms are the Rosseland optical depth across the cell along each coordinate. This equation can be written down for all the cells in which the PDA is required. In some cases, the T value will be well determined if they neighbor the diffusion region - these values act as boundary conditions. The result is a system of linear equations that can be formally solved.

\end{document}